\documentclass[aps,amsmath,amssymb,prd,reprint,superscriptaddress,nofootinbib,preprintnumbers]{revtex4-1}
\usepackage{graphicx}

\begin{document}

\preprint{TU-1110}

\title{No-go theorem of anisotropic inflation via Schwinger mechanism}

\author{Hiroyuki Kitamoto}
\email{hiroyuki.kitamoto.d3@tohoku.ac.jp}
\affiliation{Frontier Research Institute for Interdisciplinary Sciences, Tohoku University,
Aramaki Aza-Aoba 6-3, Aoba-ku, Sendai, Miyagi 980-8578, Japan}

\begin{abstract}
In the presence of a dilatonic coupling between an inflaton and a $U(1)$ gauge field, 
a persistent electric field (i.e., an anisotropic inflation) is obtained 
as a solution of the classical field equations. 
We introduce charged, massive, and conformally coupled fields into this model 
and study the pair production of charged particles. 
The semiclassical approach allows us to evaluate the induced current due to the pair production 
on the general dilatonic factor and electric field. 
Solving the field equations with the induced current, 
we find that the electric field shows a damped oscillation, 
whose amplitude decays to zero regardless of the values of the masses of charged fields. 
In other words, we derive a no-go theorem of anisotropic inflation 
by taking into account the Schwinger mechanism. 
\end{abstract}

\maketitle

\section{Introduction}\label{S-I}

In four dimensions, the electromagnetic field becomes diluted during inflation 
as long as the kinetic term of the gauge field is canonical. 
Introducing a dilatonic coupling between the inflaton and the $U(1)$ gauge field, 
Watanabe, Kanno, and Soda showed that a persistent electric field can be obtained 
as a solution of the classical field equations \cite{Watanabe2009}. 
The persistence of the electric field is equivalent to that of the anisotropic expansion rate. 
This model thus describes an anisotropic inflation at the classical level. 

It should be recalled that 
the presence of a strong electric field leads to the pair production of charged particles \cite{Schwinger1951}. 
This so-called Schwinger mechanism induces the $U(1)$ current, 
which screens the electric field at least in Minkowski space. 
We thus have a conjecture that 
the induced current screens the electric field also in the inflation model with the dilatonic coupling; 
i.e., the inflationary expansion becomes isotropic via the Schwinger mechanism.\footnote{
The Schwinger mechanism can affect not only the electric field but also the magnetic field. 
See, e.g., \cite{Kobayashi2014,Sobol2018,Kobayashi2019} for the effect on the magnetic field.} 

In the previous study \cite{Kitamoto2018}, 
we evaluated the first-order backreaction to the electric field 
and found that the electric field starts to decay with the cosmic expansion. 
The previous study indicates that the no-anisotropic hair conjecture is true. 
However, it can describe only the initial behavior of the backreaction 
because the induced current is evaluated on the classical dilatonic factor and electric field. 

In order to prove the no-anisotropic hair conjecture, 
we need to evaluate the whole time evolution of the backreaction. 
In this paper, we evaluate the induced current on the general dilatonic factor and electric field 
by using the semiclassical method proposed in \cite{Pokrovskii1961}. 
The induced current is expressed as a functional of the dilatonic factor and the electric field. 
Solving the field equations with the induced current, 
we can evaluate the backreaction on the whole time range.\footnote{
Such methods were developed to study the Schwinger mechanism in Minkowski space. 
See, e.g., \cite{Gatoff1987,Kluger1991,Kluger1992,Rau1994,Rau1995,Kluger1998,Bloch1999}.} 

The organization of this paper is as follows. 
In Sec. \ref{S-CS}, we review that a persistent electric field is obtained as a classical solution 
in the inflation model with the dilatonic coupling. 
In Sec. \ref{S-IC}, with the help of the semiclassical method, 
we study the pair production of charged particles 
without specifying the time dependences of the dilatonic factor and the electric field.  
Specifically, we derive a general expression of the induced current as a functional of them. 
In Sec. \ref{S-BR}, we evaluate the backreaction to the dilatonic factor and the electric field 
by solving the field equations with the induced current. 
We conclude with a discussion in Sec. \ref{S-C}. 

\section{Classical solution}\label{S-CS}

Here, we give a brief review of the anisotropic inflation model proposed in \cite{Watanabe2009}. 
The specific feature of this model is that the inflaton $\varphi$ is coupled to the $U(1)$ gauge field $A_\mu$
through the following dilatonic factor: 
\begin{align}
S_{\varphi,A}=\int\sqrt{-g}d^4x\big[&\frac{M_\text{pl}^2}{2}R
-\frac{1}{2}g^{\mu\nu}\partial_\mu\varphi\partial_\nu\varphi-V(\varphi) \notag\\
&-\frac{1}{4}f^2(\varphi)g^{\mu\rho}g^{\nu\sigma}F_{\mu\nu}F_{\rho\sigma}\big], 
\label{Model1}\end{align}
\begin{align}
f(\varphi)=\exp\big(\frac{2c}{M_\text{pl}^2}\int d\varphi\ \frac{V}{\partial_\varphi V}\big), 
\label{Model2}\end{align}
where $c$ is a free parameter larger than unity, $c>1$. 

Imposing the slow-roll condition, 
\begin{align}
\epsilon_V\equiv \frac{M_\text{pl}^2}{2}\big(\frac{\partial_\varphi V}{V}\big)^2\ll 1,\hspace{1em} 
\eta_V\equiv M_\text{pl}^2\frac{\partial_\varphi^2 V}{V}\ll 1, 
\end{align}
the background spacetime is approximated by de Sitter space, 
\begin{align}
ds^2\simeq-dt^2+a^2(t)d\textbf{x}^2,\hspace{1em}a(t)\simeq e^{Ht}, 
\end{align}  
where $t$ is the cosmic time and $H$ is the Hubble parameter. 
The variation of $H$ and the anisotropic expansion rate is suppressed by the slow-roll parameters. 

We consider the homogeneous background gauge field and adopt the temporal gauge: 
\begin{align}
A_0=0,\hspace{1em}A_i=A(t)\delta_i^{\ 1}. 
\label{A1}\end{align}
The physical scale of the electric field (we simply call it the electric field in this paper) is given by 
\begin{align}
E=-fa^{-1}\dot{A}, 
\label{E}\end{align}
where $\dot{}$ means the derivative with respect to the cosmic time. 

Solving the classical field equations under the slow-roll condition\footnote{
In this paper, a persistent electric field is prepared completely at the classical level. 
On the other hand, in \cite{Sobol2018,Gorbar2019,Sobol2020}, 
a persistent electric field is prepared by considering the horizon crossing mode of the gauge field 
which may experience the quantum-to-classical transition. 
We thus study the Schwinger mechanism in the situation 
different from \cite{Sobol2018,Gorbar2019,Sobol2020}.}
\begin{align}
3M_\text{pl}^2H^2\simeq V, 
\label{EQ1}\end{align}
\begin{align}
3H\dot{\varphi}+\partial_\varphi V-f^{-1}\partial_\varphi f \cdot E^2\simeq 0, 
\label{EQ2}\end{align}
\begin{align}
(a^2fE)^\cdot=0, 
\label{EQ3}\end{align}
we obtain 
\begin{align}
f=a^{-2},\hspace{1em}E=E_0\equiv \frac{\sqrt{3(c-1)\epsilon_V}}{c}M_\text{pl}H, 
\label{Sol1}\end{align}
where (\ref{EQ1}) determines the scalar factor to be approximately $e^{Ht}$.  
It should be noted that the electric field is persistent; 
i.e., its variation is suppressed by the slow-roll parameters. 

The persistent electric field gives the persistent anisotropic expansion rate: 
\begin{align}
ds^2=-dt^2+a^2(t)[&e^{-4\sigma(t)}dx_1^2 \notag\\
&+e^{2\sigma(t)}dx_2^2+e^{2\sigma(t)}dx_3^2], 
\end{align}
\begin{align}
\dot{\sigma}=\frac{E_0^2}{9M_\text{pl}^2H}=\frac{c-1}{3c^2}\epsilon_V H. 
\end{align}
Note that $\dot{\sigma}/H$ is suppressed by the slow-roll parameter. 
This is why we can evaluate the anisotropic expansion rate 
as a linear response from the electric field in the isotropic inflation. 

As reviewed above, the model (\ref{Model1})--(\ref{Model2}) realizes the anisotropic inflation 
by making the electric field persistent. 
However, if charged particles are present, 
the anisotropic inflation may become unstable due to their pair production. 
In the subsequent sections, we discuss the instability of the anisotropic inflation 
via the Schwinger mechanism. 

\section{Induced current due to pair production}\label{S-IC}

As an example of test fields, 
we consider a charged, massive, conformally coupled scalar field:\footnote{
In this paper, we introduce charged test fields into the inflation model (\ref{Model1})--(\ref{Model2}). 
On the other hand, in \cite{Shakeri2019}, the Schwinger mechanism is studied in the minimal setup 
where the inflaton takes the role not only as the dilaton but also as the charged field.} 
\begin{align}
S_\phi=\int\sqrt{-g}d^4x
\big[&-g^{\mu\nu}(\partial_\mu+ieA_\mu)\phi^*(\partial_\nu-ieA_\nu)\phi \notag\\
&-(\frac{1}{6}R+m^2)\phi^*\phi\big]. 
\end{align}
Inversely solving (\ref{E}), the gauge field is expressed by the dilatonic factor and the electric field, 
\begin{align}
A(t)=-\int^t dt'af^{-1}E(t'). 
\label{A2}\end{align}
We often use abbreviations such as $a(t)f^{-1}(t)E(t) \to af^{-1}E(t)$ in this paper. 
The Schwinger mechanisms on some fixed $(f,E)$ were already studied; e.g.,  
\cite{Kobayashi2014,Hayashinaka2016,Banyeres2018} studied the $(f,E)=(1,\text{const.})$ case, 
and \cite{Geng2017,Kitamoto2018} studied the $(f,E)=(a^{-2},\text{const.})$ case. 
In both cases, $a$ was fixed to be the de Sitter one. 
Here, we study the Schwinger mechanism on general $(f,E)$ (and general $a$) 
to evaluate the whole time evolution of the backreaction. 

For convenience, we use the conformal time $\tau=\int^tdt'a^{-1}(t')$ and the conformal transformation: 
\begin{align}
\tilde{\phi}(x)=a(\tau)\phi(x).  
\end{align}
The scalar field can be expanded as follows: 
\begin{align}
\tilde{\phi}(x)=\int\frac{d^3k}{(2\pi)^3}
[a_\textbf{k}\tilde{\phi}_\textbf{k}(\tau)
+b_{-\textbf{k}}^\dagger\tilde{\phi}_\textbf{k}^*(\tau)]e^{i\textbf{k}\cdot\textbf{x}},  
\end{align}
where the annihilation and the creation operators satisfy 
\begin{align}
[a_\textbf{k},a_{\textbf{k}'}^\dagger]=[b_\textbf{k},b_{\textbf{k}'}^\dagger]
=(2\pi)^3\delta^{(3)}(\textbf{k}-\textbf{k}'), 
\end{align}
and the other commutators are zero. 
The Klein-Gordon equation is given by 
\begin{align}
\big[\frac{d^2}{d\tau^2}+\omega_\textbf{k}^2(\tau)\big]\tilde{\phi}_\textbf{k}(\tau)=0, 
\label{KG}\end{align}
\begin{align}
\omega_\textbf{k}^2(\tau)=[k_1-eA(\tau)]^2+k_\perp^2+m^2a^2(\tau), 
\label{omega}\end{align}
where $\textbf{k}=(k_1,k_2,k_3)$ is the comoving momentum and $k_\perp^2=k_2^2+k_3^2$. 

For slowly varying $\omega_\textbf{k}(\tau)$, 
we can evaluate the pair production based on the adiabatic mode function $\bar{\phi}_\textbf{k}(\tau)$: 
\begin{align}
&\tilde{\phi}_\textbf{k}(\tau)
=\alpha_\textbf{k}(\tau)\bar{\phi}_\textbf{k}(\tau) 
+\beta_\textbf{k}(\tau)\bar{\phi}_\textbf{k}^*(\tau), \notag\\
&\frac{d}{d\tau}\tilde{\phi}_\textbf{k}(\tau)
=-i\omega_\textbf{k}(\tau)\alpha_\textbf{k}(\tau)\bar{\phi}_\textbf{k}(\tau) \notag\\
&\hspace{5em}+i\omega_\textbf{k}(\tau)\beta_\textbf{k}(\tau)\bar{\phi}_\textbf{k}^*(\tau), 
\label{Bogoliubov}\end{align}
\begin{align}
\bar{\phi}_\textbf{k}(\tau)=&\frac{1}{\sqrt{2\omega_\textbf{k}(\tau)}}e^{-i\Theta_\textbf{k}(\tau)}, \notag\\
&\Theta_\textbf{k}(\tau)=\int^\tau d\tau'\omega_\textbf{k}(\tau'). 
\end{align}
The Bogoliubov coefficients $\alpha_\textbf{k}(\tau)$ and $\beta_\textbf{k}(\tau)$ satisfy 
$|\alpha_\textbf{k}(\tau)|^2-|\beta_\textbf{k}(\tau)|^2=1$ as a consequence of the commutation relation. 

The purpose in this section is to evaluate the $U(1)$ current (density): 
\begin{align}
\tilde{j}_\mu=-ie[\langle\tilde{\phi}^*(\partial_\mu-ieA_\mu)\tilde{\phi}\rangle
-\langle\tilde{\phi}(\partial_\mu+ieA_\mu)\tilde{\phi}^*\rangle]. 
\label{j1}\end{align}
From (\ref{A1}), it is written as follows: 
\begin{align}
\tilde{j}_0=0,\hspace{1em}\tilde{j}_i=\tilde{j}(\tau)\delta_i^{\ 1}. 
\end{align}
Substituting (\ref{Bogoliubov}) into (\ref{j1}), the induced current due to the pair production is given by 
\begin{align}
\tilde{j}(\tau)=\tilde{j}_\text{cond}(\tau)+\tilde{j}_\text{pol}(\tau), 
\label{j2}\end{align}
\begin{align}
\tilde{j}_\text{cond}(\tau)
=2e\int\frac{d^3k}{(2\pi)^3}\frac{k_1-eA(\tau)}{\omega_\textbf{k}(\tau)} |\beta_\textbf{k}(\tau)|^2, 
\label{j3}\end{align}
\begin{align}
\tilde{j}_\text{pol}(\tau)=2e\int&\frac{d^3k}{(2\pi)^3}\frac{k_1-eA(\tau)}{\omega_\textbf{k}(\tau)} \notag\\
&\times\text{Re}\big[\alpha_\textbf{k}(\tau)\beta_\textbf{k}^*(\tau)e^{-2i\Theta_\textbf{k}(\tau)}\big]. 
\label{j4}\end{align}
Since $|\beta_\textbf{k}(\tau)|^2$ and $[k_1-eA(\tau)]/\omega_\textbf{k}(\tau)$ mean 
the distribution function and the velocity of produced particles, respectively, 
\begin{align}
n_\textbf{k}(\tau)= |\beta_\textbf{k}(\tau)|^2,\hspace{1em}
v_\textbf{k}(\tau)= \frac{k_1-eA(\tau)}{\omega_\textbf{k}(\tau)}, 
\end{align}
we can identify (\ref{j3}) with the conductive current. 
As discussed later, (\ref{j4}) can be identified with the polarization current. 

First, we evaluate the conductive current (\ref{j3}). 
It should be noted that 
the adiabatic approximation breaks down 
in the vicinity of the turning point of the frequency in the complex time plane 
\begin{align}
\omega_\textbf{k}(\tau_*)\equiv 0, 
\end{align}
because $d\omega_\textbf{k}(\tau)/d\tau \gg \omega^2_\textbf{k}(\tau)$ 
and $d^2\omega_\textbf{k}(\tau)/d\tau^2 \gg \omega^3_\textbf{k}(\tau)$ there.  
In other words, the pair production occurs on this turning point. 
Specifically, the distribution function of produced particles is given by 
\begin{align}
n_\textbf{k} =\exp\big[4\text{Im}\ \Theta_\textbf{k}(\tau_*)\big]. 
\label{n1}\end{align}
See \cite{Pokrovskii1961} for the derivation of this formula. 

In (\ref{omega}), the $[k_1-eA(\tau)]^2$ term may be dominant 
compared with the $k_\perp^2+m^2a^2(\tau)$ term 
because it includes the growing factor $f^{-1}(\tau)$. 
We thus treat the latter term as the perturbation from the former term. 
For the turning point, the $[k_1-eA(\tau)]^2$ term determines its real part, 
while the $k_\perp^2+m^2a^2(\tau)$ term determines its imaginary part: 
\begin{align}
\tau_*=\text{Re}\ \tau_* -i\epsilon, 
\label{TP1}\end{align} 
\begin{align}
k_1-eA(\text{Re}\ \tau_*)\simeq 0, 
\label{TP2}\end{align}
\begin{align}
\epsilon \simeq \frac{\sqrt{k_\perp^2+m^2a^2(\text{Re}\ \tau_*)}}{ea^2f^{-1}|E|(\text{Re}\ \tau_*)}, 
\label{TP3}\end{align}
where $\epsilon=-\text{Im}\ \tau_*$ is taken positive for a convergence. 
We thus evaluate $\text{Im}\ \Theta_\textbf{k}(\tau_*)$ around the turning point: 
\begin{align}
\omega_\textbf{k}(\tau')\simeq 
\big[&e^2a^4f^{-2}|E|^2(\text{Re}\ \tau_*)
(\tau'-\text{Re}\ \tau_*)^2 \notag\\
&+k_\perp^2+m^2a^2(\text{Re}\ \tau_*)\big]^\frac{1}{2},  
\label{Int1}\end{align}
\begin{align}
\text{Im}\ \Theta_\textbf{k}(\tau_*)
=-\frac{\pi}{4} \frac{k_\perp^2+m^2a^2(\text{Re}\ \tau_*)}{ea^2f^{-1}|E|(\text{Re}\ \tau_*)}. 
\label{Int2}\end{align}

From (\ref{TP1})--(\ref{Int2}), (\ref{n1}) is given by 
\begin{align}
n_\textbf{k}=\exp\big[-\pi \frac{k_\perp^2+m^2a^2(\text{Re}\ \tau_*)}{ea^2f^{-1}|E|(\text{Re}\ \tau_*)}\big]. 
\label{n2}\end{align}
This distribution function does not include time dependence 
because $\text{Re}\ \tau_*$ is a function of $k_1$. 
It should be noted that the pair production occurs when $\tau$ exceeds $\text{Re}\ \tau_*$. 
We thus define the distribution function by introducing the step function\footnote{
The adoption of the step function corresponds to a kind of Markov limit. 
See \cite{Kluger1998,Bloch1999} for non-Markov effects in Minkowski space 
and \cite{Gorbar2019,Sobol2020} for those in inflation.}
\begin{align}
n_\textbf{k}(\tau)
=&\exp\big[-\pi \frac{k_\perp^2+m^2a^2(\text{Re}\ \tau_*)}{ea^2f^{-1}|E|(\text{Re}\ \tau_*)}\big] \notag\\
&\times\theta(\tau-\text{Re}\ \tau_*)\theta(\text{Re}\ \tau_*-\tau_0),  
\label{n3}\end{align}
where $\tau_0$ means the initial time. 

Substituting (\ref{n3}) into (\ref{j3}), the conductive current is given by 
\begin{align}
\tilde{j}_\text{cond}(\tau)&=2e\int\frac{d^3k}{(2\pi)^3}v_\textbf{k}(\tau)n_\textbf{k}(\tau) \notag\\
&=\frac{e^3}{4\pi^3}\int^t_{t_0}dt'a^3f^{-2}|E|^2(t')\text{sgn}(E(t')) \notag\\
&\hspace{5.5em}\times\exp\big[-\frac{\pi m^2}{ef^{-1}|E|(t')}\big]. 
\label{j5}\end{align}
In deriving (\ref{j5}), 
we approximated the velocity as $v_\textbf{k}(\tau)\simeq \text{sgn}(E(\text{Re}\ \tau_*))$. 
We performed the Gaussian integrals with respect to $k_2$ and $k_3$: 
\begin{align}
&\int dk_2dk_3\exp\big[-\frac{\pi k_\perp^2}{ea^2f^{-1}|E|(\text{Re}\ \tau_*)}\big] \notag\\ 
&\hspace{11em}=ea^2f^{-1}|E|(\text{Re}\ \tau_*). 
\end{align}
The $k_1$ integral was translated into the time integral:  
\begin{align}
\int &dk_1\theta(\tau-\text{Re}\ \tau_*)\theta(\text{Re}\ \tau_*-\tau_0) \notag\\
&=e\int^\tau_{\tau_0} d\tau' a^2f^{-1}|E|(\tau') 
=e\int^t_{t_0} dt' af^{-1}|E|(t'). 
\end{align}

As a consistency check, we show that (\ref{j5}) can reproduce the previous results, 
including their numerical coefficients. 
Substituting $(f,E)=(1,\text{const.})$ into (\ref{j5}), we obtain the result in \cite{Kobayashi2014}: 
\begin{align}
\tilde{j}_\text{cond}(\tau)\simeq\frac{e^3}{4\pi^3}&\frac{a^3(\tau)}{3H}|E|^2\text{sgn}(E) \notag\\
&\times\exp\big(-\frac{\pi m^2}{e|E|}\big). 
\end{align}
Substituting $(f,E)=(a^{-2},\text{const.})$ into (\ref{j5}), we obtain the result in \cite{Kitamoto2018}: 
\begin{align}
\tilde{j}_\text{cond}(\tau)\simeq\frac{e^3}{4\pi^3}&\frac{a^7(\tau)}{7H}|E|^2\text{sgn}(E) \notag\\
&\times\exp\big(-\frac{\pi m^2}{ea^2(\tau)|E|}\big). 
\end{align}
In both cases, $a$ was set to be $e^{Ht}$. 
We skipped the initial time dependences of these currents for simplicity. 

Second, we evaluate the other current (\ref{j4}). 
From (\ref{KG}) and (\ref{Bogoliubov}), we can derive the following relation: 
\begin{align}
\text{Re}\big[\alpha_\textbf{k}(\tau)\beta_\textbf{k}^*(\tau)e^{-2i\Theta_\textbf{k}(\tau)}\big]
=\frac{\omega_\textbf{k}(\tau)}{\frac{d}{d\tau}\omega_\textbf{k}(\tau)}\frac{d}{d\tau}n_\textbf{k}(\tau). 
\end{align}
As well as in (\ref{TP1})--(\ref{TP3}), we approximate the derivative of the frequency as follows: 
\begin{align}
\frac{d}{d\tau}\omega_\textbf{k}(\tau)\simeq \frac{k_1-eA(\tau)}{\omega_\textbf{k}(\tau)}
\cdot ea^2f^{-1}E(\tau). 
\end{align}
The current (\ref{j4}) is thus written as follows: 
\begin{align}
\tilde{j}_\text{pol}(\tau)=2e\int\frac{d^3k}{(2\pi)^3}
\frac{\omega_\textbf{k}(\tau)}{ea^2f^{-1}E(\tau)}\frac{d}{d\tau}n_\textbf{k}(\tau). 
\label{j6}\end{align}
Since $\omega_\textbf{k}(\tau)/[ea^2f^{-1}E(\tau)]$ means the distance of produced particles 
\begin{align}
l_\textbf{k}(\tau)=\frac{\omega_\textbf{k}(\tau)}{ea^2f^{-1}E(\tau)}, 
\end{align} 
we can identify (\ref{j6}) with the polarization current. 

Substituting (\ref{n3}) into (\ref{j6}), the polarization current is given by  
\begin{align}
\tilde{j}_\text{pol}(\tau)&=2e\int\frac{d^3k}{(2\pi)^3}l_\textbf{k}(\tau)\frac{d}{d\tau}n_\textbf{k}(\tau) \notag\\
&=\frac{e^\frac{5}{2}}{8\pi^3}a^3f^{-\frac{3}{2}}|E|^\frac{3}{2}(\tau)\text{sgn}(E(\tau)) \notag\\
&\hspace{3em}\times\frac{2}{\sqrt{\pi}}\Gamma\big(\frac{3}{2},\frac{\pi m^2}{ef^{-1}|E|(\tau)}\big), 
\label{j7}\end{align}
where $\Gamma(s,z)$ is the incomplete gamma function. 
When $m^2/[ef^{-1}|E|(\tau)]\ll 1$, the last factor becomes unity: 
\begin{align}
\frac{2}{\sqrt{\pi}}\Gamma\big(\frac{3}{2},\frac{\pi m^2}{ef^{-1}|E|(\tau)}\big)\simeq 1. 
\end{align}
When $m^2/[ef^{-1}|E|(\tau)]\gg 1$, it behaves as the suppression factor: 
\begin{align}
\frac{2}{\sqrt{\pi}}\Gamma\big(\frac{3}{2},\frac{\pi m^2}{ef^{-1}|E|(\tau)}\big) 
\simeq&\frac{2}{\sqrt{\pi}}\big[\frac{\pi m^2}{ef^{-1}|E|(\tau)}\big]^\frac{1}{2} \notag\\
&\times\exp\big[-\frac{\pi m^2}{ef^{-1}|E|(\tau)}\big]. 
\end{align}
In deriving (\ref{j7}), the $k_1$ integral was performed trivially: 
\begin{align}
\int dk_1\delta(\tau-\text{Re}\ \tau_*)=ea^2f^{-1}|E|(\tau). 
\end{align}
We performed the integrals with respect to $k_2$ and $k_3$: 
\begin{align}
\int &dk_2dk_3\sqrt{k_\perp^2+m^2a^2(\tau)} \notag\\
&\hspace{3em}\times\exp\big[-\pi\frac{k_\perp^2+m^2a^2(\tau)}{ea^2f^{-1}|E|(\tau)}\big] \notag\\ 
&=\frac{1}{2}e^\frac{3}{2}a^3f^{-\frac{3}{2}}|E|^\frac{3}{2}(\tau) \notag\\
&\hspace{3em}\times\frac{2}{\sqrt{\pi}}\Gamma\big(\frac{3}{2},\frac{\pi m^2}{ef^{-1}|E|(\tau)}\big). 
\end{align}

In the semiclassical description of produced particles, 
the induced current is divided into the conductive current and the polarization current. 
We expressed them as functionals of the dilatonic factor and the electric field: (\ref{j5}) and (\ref{j7}). 
In the next section, we construct the self-consistent equations for the dilatonic factor and the electric field 
by using the general expression of the induced current. 

We mention the induced current of charged Dirac fields before moving to the next section.  
A parallel study can be done for Dirac fields 
because they are conformally coupled to the background spacetime. 
The translation process from conformally coupled scalar fields to Dirac fields is quite simple; 
we have only to introduce the overall factor $2$ into (\ref{j5}) and (\ref{j7}), 
which comes from the spin sum. 
The same factor was discussed in \cite{Kluger1992,Bloch1999,Hayashinaka2016,Sobol2018}. 
Therefore, we explicitly discuss only the contribution from conformally coupled scalar fields. 

\section{Backreaction to the dilatonic factor and electric field}\label{S-BR}

Under the slow-roll condition, the field equations with the induced current are given by 
\begin{align}
3M_\text{pl}^2H^2\simeq V, 
\label{EQ1'}\end{align}
\begin{align}
3H\dot{\varphi}+\partial_\varphi V-f^{-1}\partial_\varphi f \cdot E^2\simeq 0, 
\label{EQ2'}\end{align}
\begin{align}
(a^2fE)^\cdot+a^{-1}\tilde{j}[f,E]=0, 
\label{EQ3'}\end{align}
where we substituted the general expression of the induced current derived in the previous section: 
\begin{align}
\tilde{j}[f,E]
&=\frac{e^3}{4\pi^3}\int^t_{t_0}dt'a^3f^{-2}|E|^2(t')\text{sgn}(E(t')) \notag\\
&\hspace{5.5em}\times\exp\big[-\frac{\pi m^2}{ef^{-1}|E|(t')}\big] \notag\\
&\hspace{1em}+\frac{e^\frac{5}{2}}{8\pi^3}a^3f^{-\frac{3}{2}}|E|^\frac{3}{2}(t)\text{sgn}(E(t)) \notag\\
&\hspace{4em}\times\frac{2}{\sqrt{\pi}}\Gamma\big(\frac{3}{2},\frac{\pi m^2}{ef^{-1}|E|(t)}\big),  
\label{j8}\end{align}
where the nonlocal term is the conductive current and the local term is the polarization current. 
In fact, we may express (\ref{j8}) as $\tilde{j}[a,f,E]$ 
because we do not specify the time dependence of the scale factor in deriving it. 
Note that under the slow-roll condition, 
(\ref{EQ1'}) determines the scale factor as $a(t)\simeq e^{Ht}$, 
independently of the other two equations. 
Therefore, we abbreviate $\tilde{j}[a,f,E]$ as $\tilde{j}[f,E]$. 

From (\ref{Model2}), the derivative of $f$ is given by 
\begin{align}
\partial_\varphi f=\frac{2c}{\sqrt{2\epsilon_V}M_\text{pl}}f. 
\label{repeat}\end{align} 
Using (\ref{EQ1'}) and (\ref{repeat}), we rewrite (\ref{EQ2'}) as follows: 
\begin{align}
\dot{\varphi}+\sqrt{2\epsilon_V}M_\text{pl}H=\frac{2c}{3\sqrt{2\epsilon_V}M_\text{pl}H}E^2. 
\label{EQ2''}\end{align}
Multiplying both sides by $a^{4c}f^2$ and using (\ref{repeat}), 
we furthermore rewrite it as follows: 
\begin{align}
(a^{4c}f^2)^\cdot=\frac{4c^2}{3\epsilon_V M_\text{pl}^2H}a^{4c}f^2E^2. 
\label{EQ2'''}\end{align}

On the other hand, (\ref{EQ3'}) can be integrated as follows: 
\begin{align}
a^2fE=a_0^2f_0E_0-\int^t_{t_0}dt' a^{-1}\tilde{j}(t'), 
\label{EQ3''}\end{align}
where $a_0$, $f_0$, and $E_0$ are the initial values 
of the scale factor, the dilatonic factor, and the electric field, respectively. 
Substituting (\ref{EQ3''}) into (\ref{EQ2'''}) and integrating it, we obtain 
\begin{align}
a^{4c}f^2=\frac{4c^2}{3\epsilon_V M_\text{pl}^2H}&\int dt a^{4(c-1)} \notag\\
&\times\big[a_0^2f_0E_0-\int^t_{t_0}dt' a^{-1}\tilde{j}(t')\big]^2, 
\label{EQ4'}\end{align}
where the first integral is an indefinite integral. 

Let us consider the classical limit where any charged field is absent. 
The indefinite integral in (\ref{EQ4'}) is evaluated as follows: 
\begin{align}
a^{4c}f^2=\frac{c^2}{3(c-1)\epsilon_V M_\text{pl}^2H^2}a^{4(c-1)}a_0^4f_0^2E_0^2, 
\label{EQ4}\end{align}
where we neglected the integration constant term. 
This term is subdominant compared with the $a^{4(c-1)}$ term because $c>1$. 
From (\ref{EQ4}) and the classical limit of (\ref{EQ3''}), we obtain 
\begin{align}
f=a^{-2},\hspace{1em}E=E_0=\frac{\sqrt{3(c-1)\epsilon_V}}{c}M_\text{pl}H, 
\label{CS}\end{align}
where we normalized the initial values of the scale factor and the dilatonic factor: 
$a_0=1$ (i.e., $t_0=0$), $f_0=1$. 

We identified the initial value of the electric field by considering the classical limit. 
Let us go back to the case where a charged field is present. 
Using the explicit value of $E_0$ in (\ref{CS}), we can express (\ref{EQ2'''}) as follows: 
\begin{align}
(a^{4c}f^2)^\cdot=4(c-1)Ha^{4c}f^2E^2/E_0^2. 
\label{EQ2''''}\end{align}
Consequently, what we need to solve are the self-consistent equations 
for the dilatonic factor and the electric field: (\ref{EQ3'}) with (\ref{j8}) and (\ref{EQ2''''}). 
For the scale factor, we may substitute $a=e^{Ht}$ into these two equations under the slow-roll condition. 

\begin{figure}[tbp]
\begin{minipage}{\hsize}
\begin{center}
\includegraphics[width=8cm]{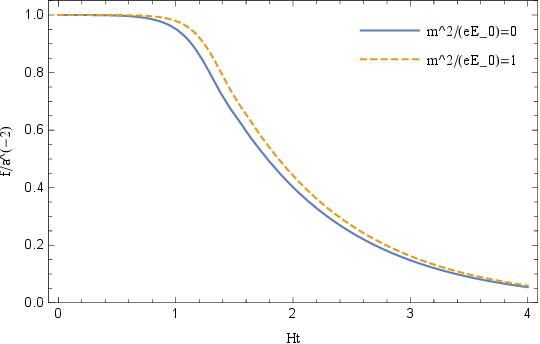}
\end{center}
\end{minipage}\\
\vspace{1em}
\begin{minipage}{\hsize}
\begin{center}
\includegraphics[width=8cm]{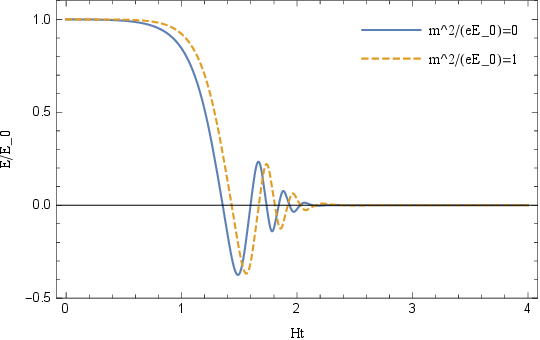}
\end{center}
\end{minipage}
\caption{
Taking the coupling and the free parameter in (\ref{Model2}) as $e=0.1$, $c=1.5$, 
the time evolutions of $f$ and $E$ were calculated.
From (\ref{CS}), $E_0/H^2$ is expressed as $E_0/H^2=\sqrt{\frac{3(c-1)}{c}}/\sqrt{8\pi^2 A_s}$, 
where $A_s$ is the scalar amplitude $A_s=H^2/(8\pi^2\epsilon_H M_\text{pl}^2)$, 
$\epsilon_H\equiv-\dot{H}/H^2=\epsilon_V/c$. 
We determined it from the observed value $A_s=2\times 10^{-9}$.}
\label{e01}\end{figure}

\begin{figure}[tbp]
\begin{minipage}{\hsize}
\begin{center}
\includegraphics[width=8cm]{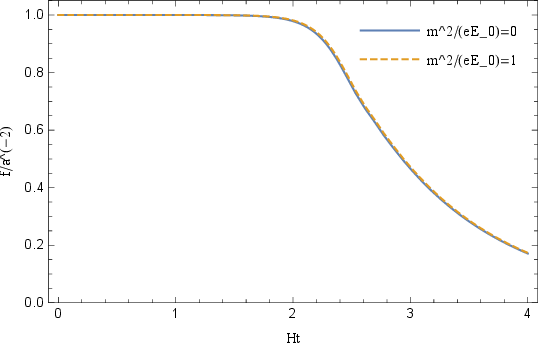}
\end{center}
\end{minipage}\\
\vspace{1em}
\begin{minipage}{\hsize}
\begin{center}
\includegraphics[width=8cm]{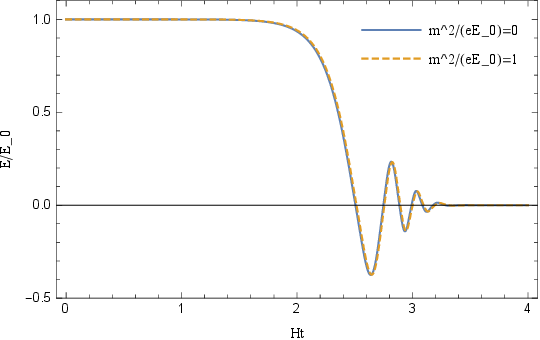}
\end{center}
\end{minipage}
\caption{Taking the coupling as $e=0.01$, 
the time evolutions of $f$ and $E$ were calculated. 
The other conditions were set to be the same as Fig. \ref{e01}. 
The coupling is $1/10$ times smaller than that in Fig. \ref{e01}, 
while the plateau range (i.e., the classical range) is only a few times longer than that in Fig. \ref{e01}. 
This is because the dilatonic factor enhances the coupling as $ef^{-1}$. 
Compared with Fig. \ref{e01}, the mass difference becomes smaller during the coupling enhancement.} 
\label{e001}\end{figure}

We show the numerical solutions of (\ref{EQ3'}) with (\ref{j8}) and (\ref{EQ2''''}) 
in Figs. \ref{e01} and \ref{e001}. 
As seen in these figures, 
the decay of the dilatonic factor $f$ eventually becomes faster than $a^{-2}$ 
due to the backreaction. 
From (\ref{EQ2''''}), we can identify the eventual behavior as $f\propto a^{-2c}$. 
Furthermore, these figures show that the electric field $E$ shows a damped oscillation, 
whose amplitude decays to zero even though $m^2/(eE_0)=1$. 
This is a specific feature of the Schwinger mechanism 
in the inflation model (\ref{Model1})--(\ref{Model2}). 

From the general expression of the induced current (\ref{j8}), 
we can see that the suppression factor of the Schwinger mechanism is given by 
\begin{align}
\exp\big(-\frac{\pi m^2}{ef^{-1}|E|}\big). 
\end{align}
The growing $f^{-1}$ makes the exponent diluted, and thus the suppression factor does not work. 
The nonzero initial value of the exponent can just delay the onset of the Schwinger mechanism. 
This is why the Schwinger mechanism completely screens the electric field 
regardless of the values of the masses of charged fields. 

\begin{figure}[tbp]
\begin{center}
\includegraphics[width=8cm]{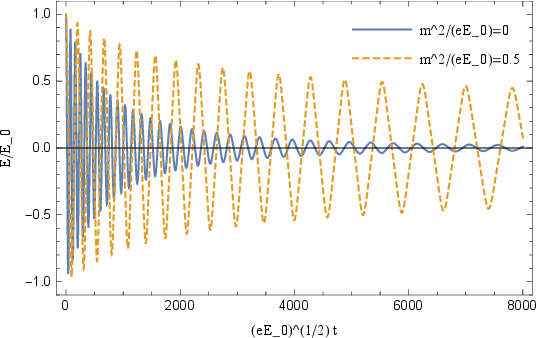}
\end{center}
\caption{Taking the coupling as $e=1$, the time evolution of $E$ was calculated. 
The $E_0$ dependence appears only in $m^2/(eE_0)$ 
after rescaling the horizontal and the vertical axes.}
\label{e1-a1}\end{figure}

For comparison, we mention the Schwinger mechanism in Minkowski space. 
In the flat limit where $a=1$ and $f=1$, (\ref{EQ3'}) with (\ref{j8}) reduces to 
\begin{align}
\dot{E}+j^\text{flat}[E]=0, 
\end{align}
\begin{align}
j^\text{flat}[E]
&=\frac{e^3}{4\pi^3}\int^t_{t_0}dt'|E|^2(t')\text{sgn}(E(t')) 
\exp\big[-\frac{\pi m^2}{e|E|(t')}\big] \notag\\
&\hspace{1em}+\frac{e^\frac{5}{2}}{8\pi^3}|E|^\frac{3}{2}(t)\text{sgn}(E(t)) 
\frac{2}{\sqrt{\pi}}\Gamma\big(\frac{3}{2},\frac{\pi m^2}{e|E|(t)}\big). 
\label{j9}\end{align}
This self-consistent equation for the electric field is what we need to solve. 
We show the numerical solutions in Fig. \ref{e1-a1}. 
As seen in this figure, $E$ shows a damped oscillation 
whose amplitude approaches a finite value 
if the masses of charged fields are finite. 
This is because the suppression factor does not include the $f$ dependence 
in Minkowski space. 

\section{Conclusion}\label{S-C}

Employing the semiclassical method, 
we studied the pair production of charged particles 
without specifying the time dependences of the dilatonic factor and the electric field (and the scale factor). 
Specifically, we derived the general expression of the induced current (\ref{j8}) 
as a functional of these backgrounds. 

We thus obtained the self-consistent equations for these backgrounds: 
(\ref{EQ1'})--(\ref{EQ3'}) with (\ref{j8}).  
Solving them numerically, 
we evaluated the whole time evolutions of the dilatonic factor and the electric field 
including the backreaction from the Schwinger mechanism. 

We found that the decay of the dilatonic factor $f$ becomes faster than the classical one. 
From (\ref{EQ2''''}), the eventual behavior of $f$ is given by $f\propto a^{-2c}$. 
Furthermore, we found that the electric field shows a damped oscillation, 
whose amplitude decays to zero regardless of the values of the masses of charged fields. 

This is because in the presence of the dilatonic factor, 
the suppression factor of the Schwinger mechanism is given 
not by $\exp[-\pi m^2/(e|E|)]$, but by $\exp[-\pi m^2/(ef^{-1}|E|)]$. 
Since the growing $f^{-1}$ makes the exponent diluted, 
the suppression factor can just delay the onset of the Schwinger mechanism. 
The Schwinger mechanism eventually cancels out the electric field as well as in the massless case. 
We thus conclude that as long as charged and conformally coupled fields are present,  
the no-go theorem of anisotropic inflation holds true regardless of the values of their masses.\footnote{
Here, we consider the case where 
the total duration of inflation is sufficiently longer than the minimum value, about 60 $e$-folds. 
Looking at Figs. \ref{e01} and \ref{e001}, 
it only has to be more than several $e$-folds longer than the minimum value. 
In contrast, if it was just about 60 $e$-folds as the Swampland conjectures suggest \cite{Agrawal2018}, 
this model could leave an observable anisotropic hair.} 

It should be noted that we considered conformally coupled fields in this paper. 
The numerical solutions show that the electric field eventually exceeds zero. 
If supercurvature modes are present, 
their contribution becomes dominant in the weak electric field region, 
and it cannot be evaluated by the semiclassical method \cite{Kobayashi2014,Banyeres2018}. 
This is why we imposed the conformally coupled condition. 
It is a future subject to study the contribution from supercurvature modes. 

Finally, we mention that a two-form field also gives rise to an anisotropic inflation at the classical level 
if it is coupled to the inflaton through a dilatonic factor \cite{Ohashi2013}. 
It is an interesting question whether there exists a certain microscopic mechanism, 
which screens the anisotropic hair originating from the two-form field. 

\begin{acknowledgments}
This work was supported by Leading Initiative for Excellent Young Researchers, MEXT, Japan. 
We thank C. Chu, M. Hotta, K. Itakura, N. Kitajima, K. Shimada, F. Takahashi, M. Yamada, 
and K. Yonekura for discussions. 
\end{acknowledgments}

\end{document}